\documentclass{revtex4}
\input epsf
\usepackage{amsmath,amssymb} 
\usepackage{graphicx}
\begin{document}
\bibliographystyle{revtex}

\title{Higgs Boson Production Rates in Hadronic Collisions}

\author{S. ~Dawson}
\email[]{dawson@bnl.gov}
\affiliation{Physics Department, Brookhaven National Laboratory\\
Upton, N.Y.~11973}

\date{\today}

\begin{abstract}
Higgs boson production rates at hadron colliders are reviewed
with particular emphasis on progress in the calculation of higher
order QCD effects.  Emphasis is placed on the uncertainties in
the predictions for Higgs boson production.  A firm understanding
of these uncertainties is crucial for extracting new physics
signals.

\end{abstract}
\maketitle

\section{Introduction}
The search for the Higgs boson is one of the  fundamental goals
of the Tevatron and the LHC.  At the Tevatron, discovery relies on the
observation of the Higgs signal in many channels, since the
rate is quite small.  In order to disentangle
a Higgs boson  from the background,
 it is crucial to have reliable predictions for both
the signal  and the background
 and to well understand the uncertainties in the theoretical predictions.
 The
case is somewhat different at the LHC, since the Higgs boson production
rate is significantly larger than at the Tevatron.
  Here discovery is more straightforward,
and the question becomes what can we learn about the underlying physics.
This again requires a firm understanding of the production rates
and decay patterns.

In this paper, we review the predictions for the production of  the
Higgs boson in the Standard Model at the Tevatron
and at the LHC.  The current status of QCD radiative
corrections to Higgs boson production
is examined critically in order to make estimates of the
uncertainties involved in the predictions.  
The next-to-leading order (NLO) results are now available for
all of the dominant  production channels and the forefront of
activity has moved to the calculation of
  next-to-next-to-leading order (NNLO) corrections 
and the resummation of leading and next-to-leading order logarithmic
effects.  The implementation of these higher order effects in
Monte Carlo programs and the comparison with fixed order perturbative
calculations remains an active area of investigation.

\section{LHC}

Many studies have been made of the
capability of the  LHC to observe the Higgs
boson in a variety of channels.  With $30~fb^{-1}$, a Standard Model Higgs
boson will be observable at the $5\sigma$ level over the entire mass range,
$100~GeV < M_h < 1~TeV$.  With $100~fb^{-1}$, the Higgs boson will be
observable in at least $2$ channels in the same mass range.
By combining various channels, some measurements of Higgs couplings
will then be possible.  The
conclusions of these studies,
 however, typically  rely on the use of the lowest order cross
sections only.\cite{atlastdr}
  The NLO QCD results exist for all the relevant 
Higgs production and decay channels, 
but not for many of the backgrounds.

\subsection{Gluon Fusion}

\begin{figure}[t]
\begin{center}
\vspace*{.8cm}
\hspace*{-1.5cm}
\epsfxsize=9cm \epsfbox{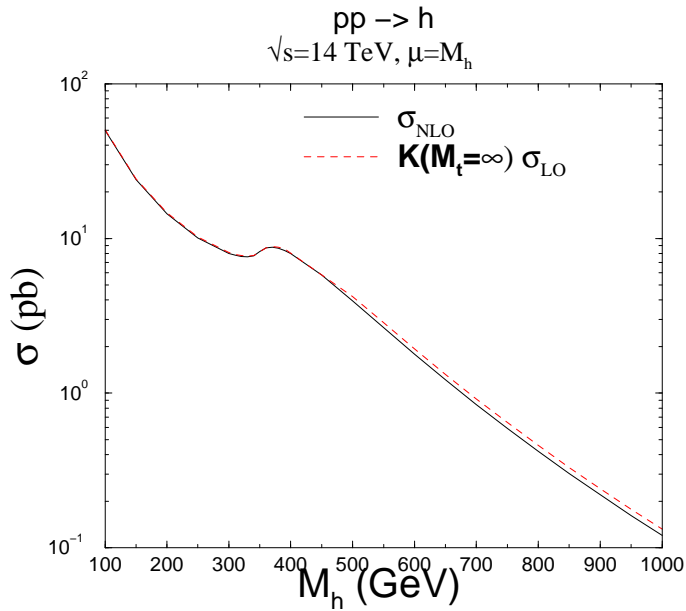}
\caption[ ]{Complete NLO result for inclusive Higgs boson production at the
LHC, $\sqrt{s}=14~TeV$,  with a renormalization/factorization scale $\mu=M_h$
 (solid).
The dashed line is obtained by combining the complete lowest order result
with the $K$ factor computed in the limit $M_t \rightarrow \infty$.}
\label{fg:signlo}
\end{center}
\end{figure}

The dominant production mechanism for the Higgs boson at the LHC is 
gluon fusion.  The NLO QCD corrections are well known, both in the $M_t
\rightarrow \infty$ limit and with the  inclusion of
 the complete $M_t$ dependence of
the result.\cite{spd}  
A convenient parameterization of the results is given by the $K$ factor,
$$
K(\mu^2)\equiv{\sigma_{NLO}(\mu^2)\over \sigma_{LO}(\mu^2)}
\quad .
$$
We have explicitely included the dependence on $\mu$ to 
emphasize the fact that the $K$ factor is typically quite
sensitive to  
the renormalization/factorization scale $\mu$.
The $M_t\rightarrow \infty$ limit provides an extremely
accurate description of the full rate at NLO, as can be seen in Fig. \ref{fg:signlo}.
By including the exact $M_h$ and $M_t$ dependence in the lowest order
result and multiplying by the $K$ factor computed in the $M_t\rightarrow
\infty$ limit, ($\sigma_{m_t\rightarrow \infty}$),
 the resulting approximation to the next-to-leading order rate
is extremely accurate all the way up to $M_h\sim 1~TeV$.  Table 1
shows the dependence of the NLO cross section on various input
parameters.  The exact NLO calculation has a small dependence on
$m_b$ through the $b$ quark loop.

\begin{table}[t]
\begin{center}
\caption{$pp\to h$ at $\sqrt{s}=14$ TeV (CTEQ5, set4; $\mu=M_h$)
\label{tabone}} 
\begin{tabular}{|l||c|c||c|c||c|c|}
\hline
& \multicolumn{2}{c||}{$M_h=100$ GeV} & \multicolumn{2}{c||}{$M_h=500$ GeV}
& \multicolumn{2}{c|}{$M_h=900$ GeV} \\
\hline
& $\sigma_{\rm NLO} (pb)$ & $\sigma^\infty_{\rm NLO} (pb)$ &
$\sigma_{\rm NLO} (pb)$ & $\sigma^\infty_{\rm NLO} (pb)$ & $\sigma_{\rm
NLO} (pb)$ & $\sigma^\infty_{\rm NLO} (pb)$ \\ 
\hline\hline
$M_b=4$ GeV & 48.3 & 48.6 & 4.1 & 4.3 & .23 & .25 \\
$M_t=175$ GeV & & & & & & \\
$\alpha_s(M_Z)=.118$ & & & & & & \\
\hline
$M_b=5$ GeV & 48.0 & 48.2 & 4.1 & 4.4 & .23 & .25 \\
$M_t=175$ GeV & & & & & & \\
$\alpha_s(M_Z)=.118$ & & & & & & \\
\hline
$M_b=4$ GeV & 50.2 & 51.0 & 4.2 & 4.5 & .24 & .27 \\
$M_t=175$ GeV & & & & & & \\
$\alpha_s(M_Z)=.120$ & & & & & & \\
\hline
$M_b=4$ GeV & 48.1 & 48.5 & 4.2 & 4.5 & .25 & .27 \\
$M_t=180$ GeV & & & & & & \\
$\alpha_s(M_Z)=.118$ & & & & & & \\
\hline
\end{tabular}
\end{center}
\end{table}

The NLO corrections to the LO rate are quite large, increasing the
cross section by about a factor of 2.  In addition, the scale dependence
remains significant.
  The LHS of Fig. 2 compares the complete LO order result with
the NLO result.  The bands represent a variation of the renormalization
scale from ${M_h\over 2} < \mu < 2 M_h$.  Note that there is no overlap
between the bands labelled LO and NLO and so, in this case, the variation
of the scale $\mu$ appears to be a poor indicator of the uncertainty of the
result.  

The accuracy of the NLO $K$ factor computed  in the $M_t
\rightarrow \infty$ limit has
encouraged two groups to undertake the calculation of the NNLO
contribution to inclusive Higgs production in this limit.\cite{hk, catani}
Using an effective theory corresponding to an 
infinite top quark mass, 
 the NNLO virtual corrections reduce
to 2-loop Feynman diagrams,
instead of the 3-loop diagrams they would be in the complete $M_t$
dependent calculation. These virtual contributions to the NNLO rate have
been computed by Harlander.\cite{harlander}

  At present,
 the existing NNLO results for the inclusive Higgs production
rate are
incomplete and make an  assumption which
  the authors
term the ``soft approximation''.  This approximation includes the leading
terms as $z\equiv M_h^2/{\hat s}\rightarrow 1$, (where
$\sqrt{{\hat s}}$ is the gluon-gluon center of
mass energy).  These leading $z\rightarrow 1$ contributions
  are of the form,
\begin{equation}
\delta(1-z),~~~~\biggl({\log^i(1-z)\over 1-z}\biggr)_+, ~~i=1,2,3\quad .
\end{equation}
These terms are expected to provide the bulk of the NNLO corrections.
The validity of the soft approximation can be tested at NLO by comparison
with the complete calculation.   Inclusion of the leading soft terms
of Eq. (1) with $i=1$, (the curve labelled NLO-SV of Fig. 2), shows that 
the soft plus virtual contributions alone underestimate
the exact NLO result by $\sim 15-20\%$.

\begin{figure}[t]
\begin{center}
\vspace*{.8cm}
\hspace*{-1.5cm}
\epsfxsize=11cm \epsfbox{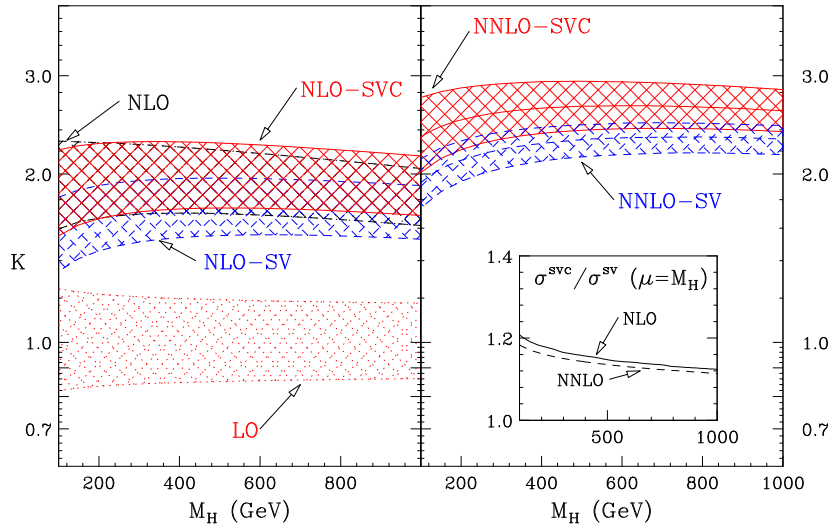}
\caption[ ]{The curved labelled LO (NLO) on the LHS are the lowest order
(next-to-leading-order)  result
for $pp\rightarrow h$ at the LHC with the renormalization scale varied
from ${M_h\over 2}< \mu <2 M_h$. The curve labelled NLO-SVC on the LHS
includes the terms of Eq. 1 with $i=j=1$, along with the NLO virtual
corrections.
 The curve labelled NNLO-SVC on the RHS
includes the terms of Eq. 1 with $i=1,2,3$ and of Eq. 2 with $j=3$,
plus the NNLO virtual corrections.
From Ref. [4].   }
\label{nnlo_cat}
\end{center}
\end{figure}

In order to obtain a more accurate approximation to the complete
rate,
the sub-leading collinear contributions can also be included.  These
terms are of the form 
\begin{equation}
\log^j(1-z),\qquad j=1,2,3.
\end{equation}
The leading collinear contributions at each order 
($j$=1 for NLO and  $j=3$ for NNLO) have been found in Refs. \cite{hk, 
catani}.
In addition, the sub-leading collinear contributions ($j=1,2$ at
NNLO) can  
be estimated from the 
resummation calculation of Ref. \cite{kls}.  From Fig. \ref{nnlo_cat},
 we see that including the collinear
$\log(1-z)$ contribution, along with the
virtual contribution and the soft terms of Eq. 1
(with $i=1$), provides an excellent
 approximation to the full
 NLO result (the curve labelled NLO-SVC in Fig. 2).  

Using
the soft plus collinear approximation 
 to the NNLO result, $i=1,2,3$ in Eq. (1) and $j=3$
in Eq. (2), yields the results shown on
the RHS of Fig. \ref{nnlo_cat} (labelled NNLO-SVC).
  We see that the NNLO corrections
 are large,
leading to a $K$ factor between $2.5$ and $3$.  The bands correspond
to varying the renormalization scale between $M_h/2 < \mu < 2 M_h$.
The scale dependence is only slightly reduced from that of the NLO
result.

Harlander and Kilgore\cite{hk}
 included also the sub-leading collinear terms
of the form $\log^2(1-z)$ and $\log(1-z)$, to
obtain the solid curves shown in
Fig. \ref{kfac_hk}.   The differences between  the 3 upper curves
in Fig 3 is due taking different approximations
for the unknown sub-leading collinear contributions
to the NNLO result
 and can be interpreted as an estimate of the uncertainty of
the result.
As is clear from this figure, the collinear contribution from the
$\log^j(1-z)$ terms is numerically quite large, (since the dotted
curve labelled ``soft'' omits the $\log^j(1-z)$ contributions).
The inset in the RHS of Fig. 2 also show the importance of the
collinear contributions.

\newdimen\figwid
\figwid=\columnwidth
\multiply\figwid by 4
\divide\figwid by 9
\begin{figure*}
\includegraphics[width=\figwid]{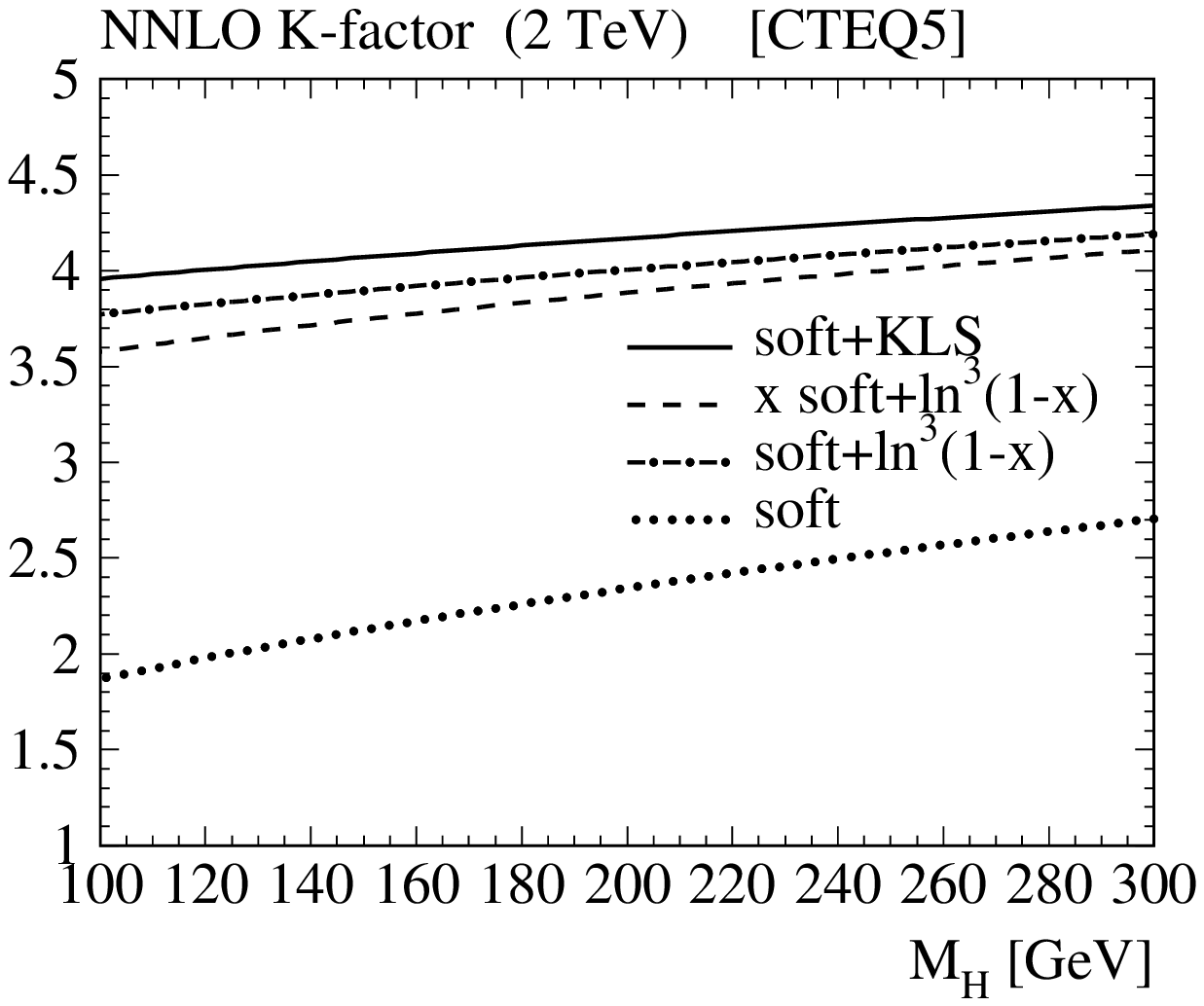}%
\hfil
\includegraphics[width=\figwid]{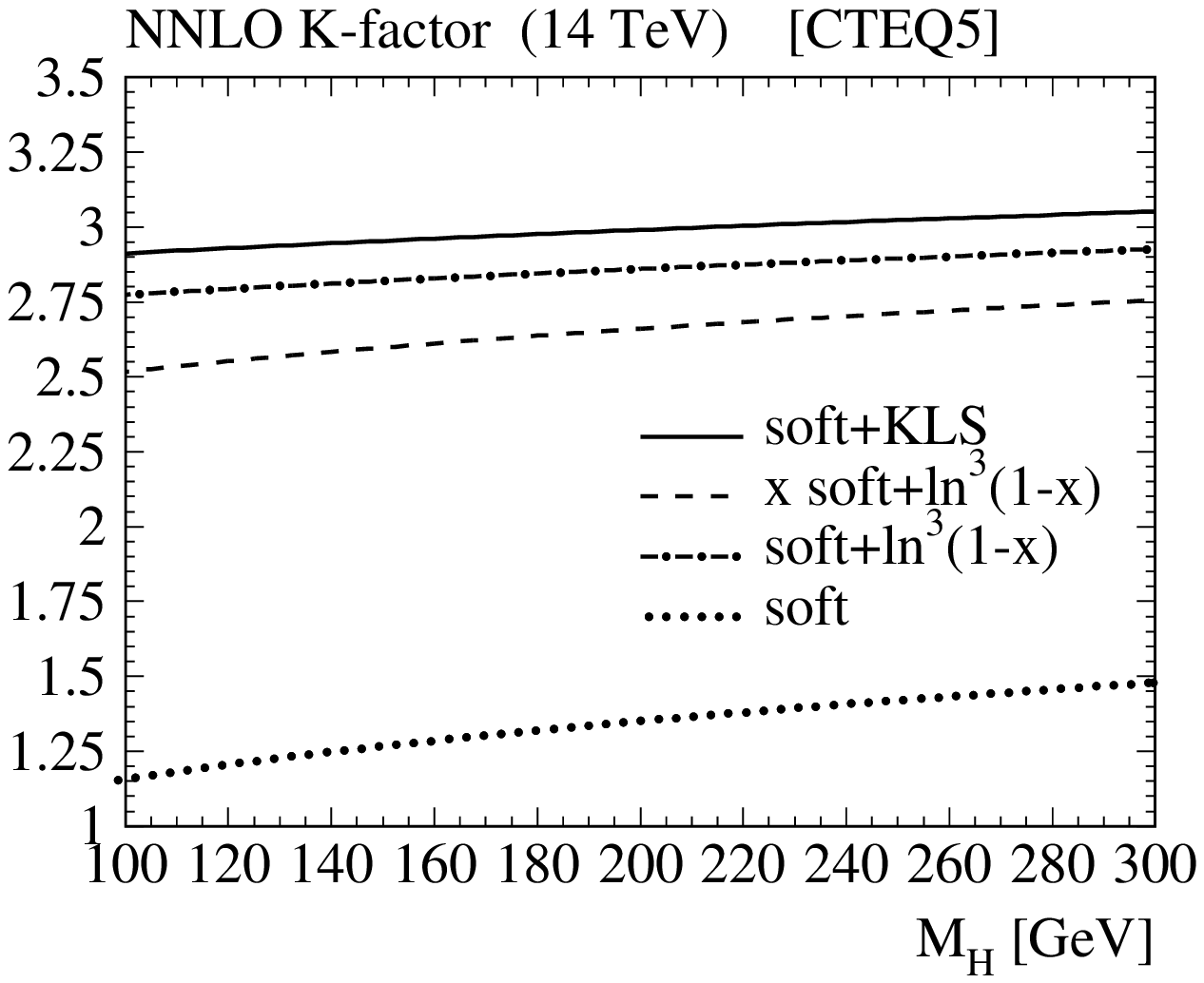}%
\caption{(a) NNLO result for the $K$ factor for $ p {\overline p}
\rightarrow h$ at $\sqrt{s}=2~TeV$ with $\mu=M_h$. (b)
NNLO result for the $K$ factor for $ p  p
\rightarrow h$ at $\sqrt{s}=14~TeV$ with $\mu=M_h$. From Ref [3].}
\label{kfac_hk}
\end{figure*}

Another important issue is the question of NNLO parton distribution
functions (pdfs).  At present only partial NNLO pdfs exist.\cite{martin}
Catani ${\it et. al.}$ use the NNLO pdfs of Ref. \cite{martin},
while Harlander and Kilgore utilize CTEQ5 NLO pdfs.
  The inclusion of NNLO
pdfs (instead of NLO pdfs)
 decreases the rate by roughly $8\%$.\cite{catani}
Clearly a complete NNLO calculation with complete NNLO pdfs
 is needed
before we can begin to extract precision results.  At present, the best
estimate is that there is still approximately a $35\%$ uncertainty in
the prediction due to scale dependence, unknown NNLO terms, incomplete
knowledge of the NNLO pdfs, and our knowledge of $\alpha_s$.

\begin{figure}[t]
\begin{center}
\vspace*{.8cm}
\hspace*{-1.5cm}
\epsfxsize=9cm \epsfbox{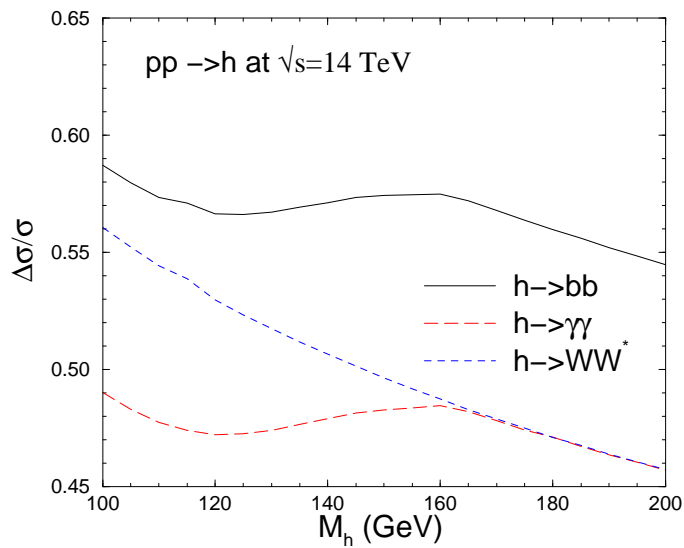}
\caption[ ]{Uncertainty on the production cross section times
branching ratio to NLO at the LHC for various
Higgs channels. $\mu$ is varied from $M_h/2 < \mu < 2 M_h$, 
$\alpha_s(M_Z)$ from 
 $.115< \alpha_s(M_Z)< .123$
and $m_b$ from
$4.5~GeV <m_b(m_b)< 5~GeV$. }
\label{sigun}
\end{center}
\end{figure}

\begin{figure}[t]
\begin{center}
\vspace*{.8cm}
\hspace*{-1.5cm}
\epsfxsize=10cm \epsfbox{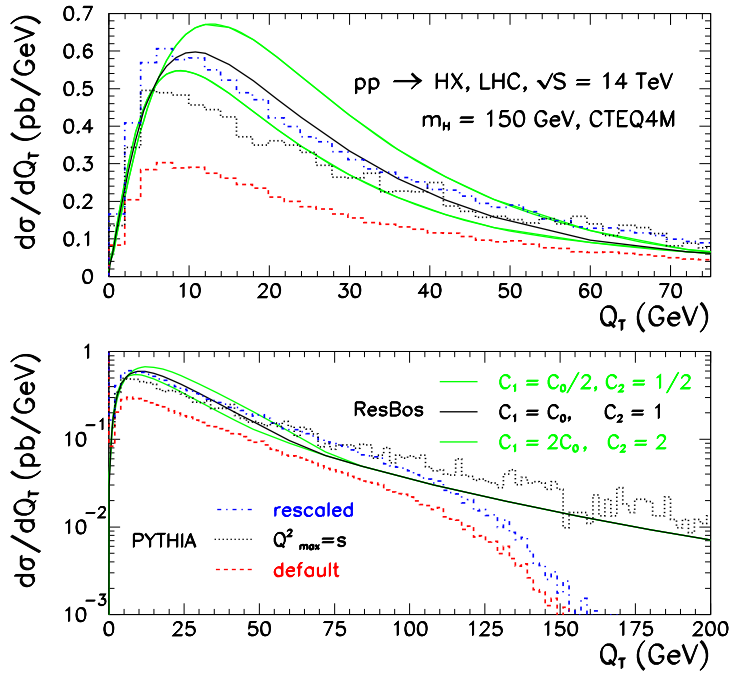}
\caption[ ]{Comparison of the Higgs $p_T$ spectrum derived
 using soft gluon 
resummation at low $p_T$, matched with the exact calculation
to ${\cal O}(\alpha_s^3)$ at
high $p_T$ (solid),  with PYTHIA (dashed). The 3 solid curves are
estimates of the uncertainty from unknown NNLO contributions.
 From Ref. [8]}
\label{ptdis}
\end{center}
\end{figure}

An important outcome of higher order calculations 
is the Higgs boson $p_T$ spectrum.  At lowest order, the Higgs
boson  is
produced with no transverse momentum. At higher orders in $\alpha_s$,
   the effects of 
soft and soft plus collinear gluon emission from the initial state 
partons are  numerically significant.
At low $p_T$, the usual factorization approximation  fails and
large logarithms of the form $\alpha_s^n \log^m(M_h^2/
p_T^2)$ appear.  These large logarithms can be resummed to
give a result which is valid at low $p_T$.
 At an intermediate value of $p_T$, the resummed form can
be matched with the exact matrix element calculation to ${\cal O}(\alpha_s^3)$, 
valid at large $p_T$.  This
result is shown in Fig. \ref{ptdis}. 
 The three solid curves in this plot represent
 an attempt to estimate uncalculated
NNLO contributions to the resummed result.  Ref. \cite{ptdis}
estimates a $\pm 10\%$ uncertainty due to these unknown  terms.  
When the complete NNLO calculation is available, it will
be possible to remove much of this uncertainty.

 Since 
experimental searches rely strongly on Monte Carlo programs, it is
important to understand how soft gluon emission is included in these
programs.
Monte Carlo  programs typically produce
the Higgs $p_T$ spectrum using parton showering,  which  correctly reproduces
the spectrum at low $p_T$, but underestimates the rate at higher $p_T$,
as can be seen in Fig. \ref{ptdis}.  The correct spectrum at high $p_T$
(as determined from the exact matrix element calculation) can
be obtained using PYTHIA by judiciously adjusting the
arbitrary renormalization scale.\cite{ptdis}

The NLO rate for Higgs plus 1- jet production
 at the LHC has been computed in
the $M_t\rightarrow \infty$ limit.\cite{kunszt}  The corrections
increase the rate by a factor of $1.5-1.6$
and are almost constant over a large
range of $M_h$, rapidity, and $p_T$.
  As with inclusive Higgs production, the
renormalization/factorization scale
uncertainty is significantly reduced by the inclusion of the NLO
contributions, although the residual uncertainty is still rather
large, $\sim \pm 20\%$.The NLO QCD results for all of the 2- and
 3- body Higgs decays have existed for
some time and are conveniently implemented in the FORTRAN code 
HDECAY.\cite{spira}
Fig. \ref{sigun} shows the variation of the inclusive
Higgs production cross section calculated to NLO multiplied
by the NLO branching ratios to various channels
 as the renomalization
scale, $\mu$, is varied from $M_h/2 < \mu < 2 M_h$, $\alpha_s(M_Z)$ is 
varied within the LEP 1-$\sigma$ limit, $.115< \alpha_s(M_Z)< .123$
and $m_b$ is varied within the particle data group range, 
$4.5~GeV <m_b(m_b)< 5~GeV$. 
The dominant source of uncertainty 
in the results of Fig. 4 is the renormalization/factorization
scale dependence.
 By measuring combinations of final states,
the uncertainty on the predictions can, however,
 be significantly reduced.\cite{zep}
The theoretical uncertainty on the branching ratios alone is 
considerably smaller than on the product $\sigma B$.\cite{bat}

\subsection{Vector Boson Fusion}
Vector boson fusion can be used to measure the $WWh$ and $ZZh$
couplings at the LHC.\cite{zep}  The NLO QCD
 corrections to $pp\rightarrow $2- jets +$h$ through
the vector boson sub-process are quite small, \cite{hvw}
and the uncertainly on the production rate
 is estimated from the small scale dependence
to be
 $\sim 1-2\%$.

It is necessary to separate the signal, $q q \rightarrow q q h$ ( which
probes the $WWh$ and $ZZh$ couplings), from
the background,  $gg\rightarrow gg h$ (which depends
only on the $t{\overline t}h$
 Yukawa coupling) and first enters at 1-loop.\cite{wbf}  Since
the dominant contribution to the background arises from  gluons in the
initial state, the background
 is enhanced by the large gluon luminosity at the LHC.
The  $gg\rightarrow gg h$ contribution to
the Higgs plus 2- jet signal has been computed 
both in the $M_t\rightarrow \infty$ limit and retaining the full $M_t$ mass
dependence and the results are shown in Fig. \ref{fig_vbf}. The
$m_t\rightarrow \infty$ limit is a good approximation to the full
result only for $M_h < 2 M_t$.   The weak boson fusion process produces
well separated, forward jets, while the jets from the gluon fusion
sub-process are more isotropic. 
Using cuts designed to enhance the vector boson fusion contribution,
shown in Fig 6(b), it is clear that it will
be possible to separate the gluon fusion sub-process from the vector
boson fusion contribution for $M_h < 2 M_t$.
   The weak boson fusion processes dominate over the gluon initiated
processes 
by about a factor of
3 to 1 after applying the appropriate cuts and for $M_h < 2 M_t$.

\newdimen\figwid
\figwid=\columnwidth
\multiply\figwid by 4
\divide\figwid by 9
\begin{figure*}
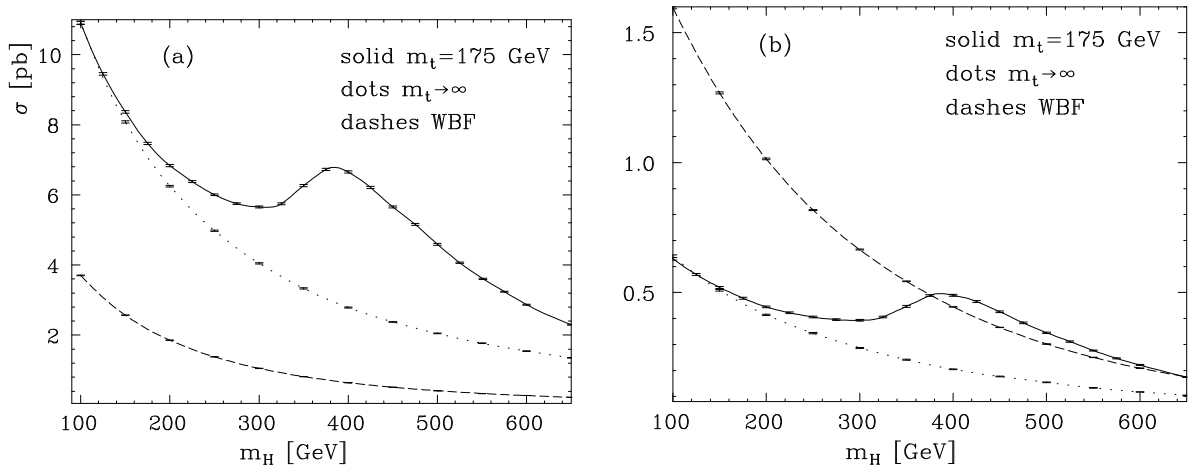

\includegraphics[width=\figwid]{P1_dawson_0706_fig6a.eps}%
\hfil
\includegraphics[width=\figwid]{P1_dawson_0706_fig6b.eps}%
\caption{(a) $pp\rightarrow {\hbox {2~jets~}}+ h$ from the weak boson
fusion sub-process at $\sqrt{s}=14~TeV$ (dashed)
and background from the $gg\rightarrow ggh$ sub-process calculated
exactly (solid) and in the $M_t\rightarrow\infty$ limit (dotted).
  (b) Same as (a), but with cuts designed to enhance the weak
boson fusion contribution.
 From Ref [14].}
\label{fig_vbf}
\end{figure*}

\subsection{Associated Production}

A Higgs boson in the mass range around $120-140~GeV$
 is particularly difficult to observe at the LHC since  the
 preferred channel, $h\rightarrow \gamma\gamma$,
suffers from  a small rate and
large backgrounds.  In this region, the associated production
channel, $ pp\rightarrow t {\overline t}h$, may be useful to confirm an
elusive Higgs signal.
  Although the production rates are small, $\sim .5-.8~pb$,
the signature with the final
 state $W^+W^- b {\overline b} b{\overline b}$ is spectacular.
This process is of particular interest since it can be used to
measure the $t {\overline t}h$ Yukawa coupling.

  The $t {\overline t}h$ process
proceeds predominantly through gluon fusion
at the LHC. The complete NLO results
have been found in Ref. \cite{ben} and are 
shown in Fig. \ref{lhc_tth}.  The NLO predictions show a significantly reduced
scale dependence and  increase the rate by roughly $20\%$ from the LO 
predictions over the entire intermediate Higgs mass range.

\newdimen\figwid
\figwid=\columnwidth
\multiply\figwid by 4
\divide\figwid by 9
\begin{figure*}
\vskip 2in
\includegraphics[width=\figwid]{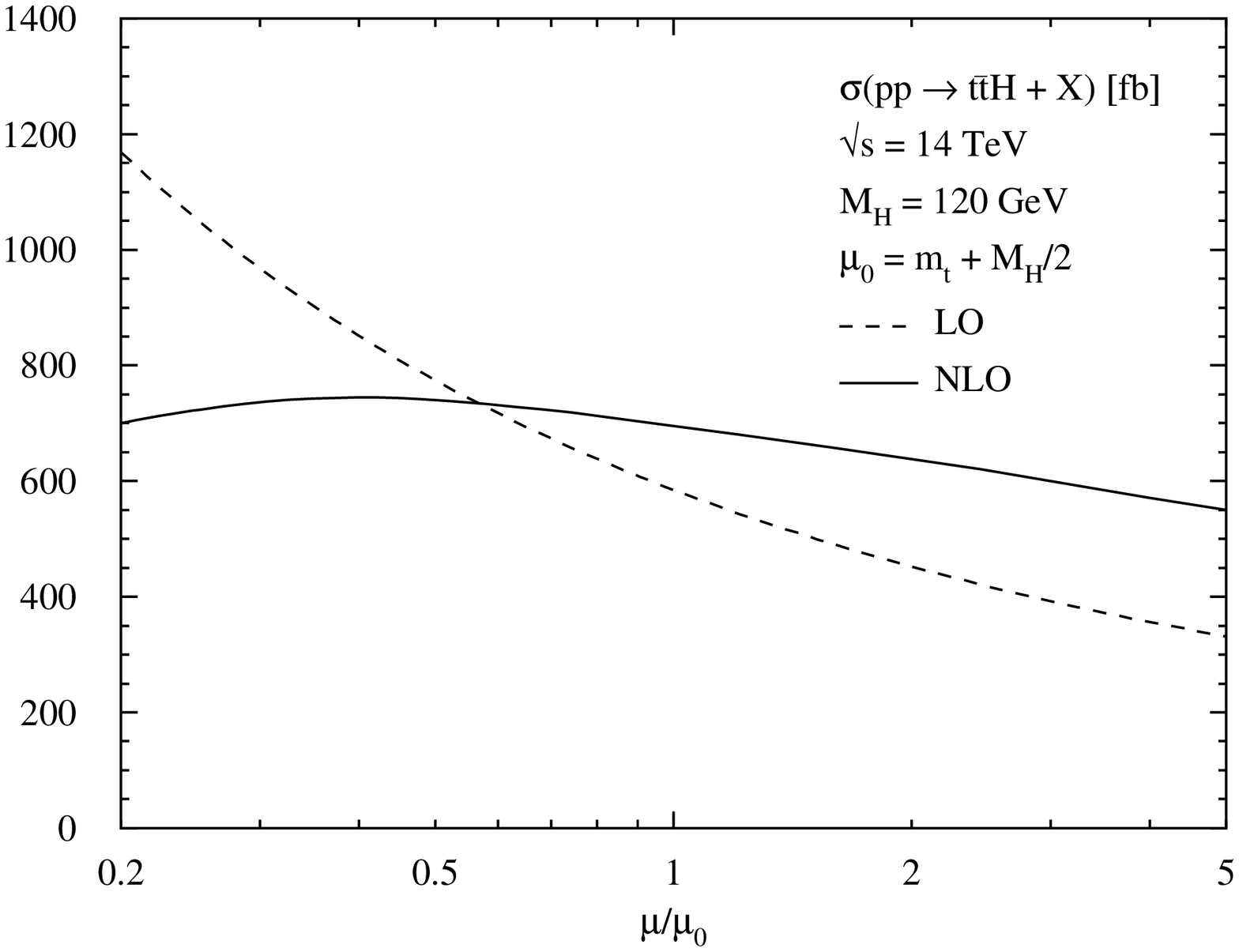}%
\hfil
\includegraphics[width=\figwid]{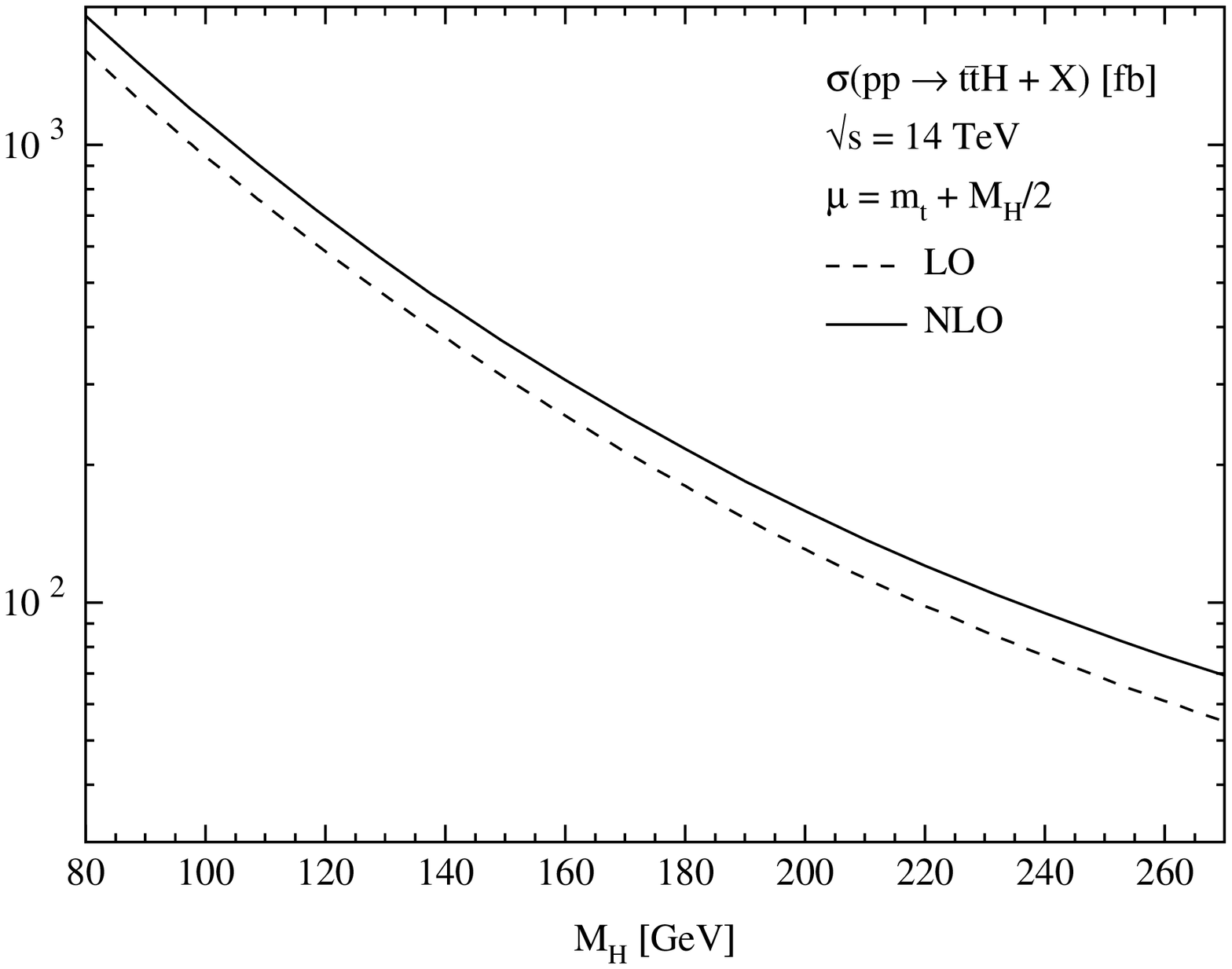}%
\vskip -1in
\caption{(a) Dependence of $\sigma_{LO, NLO}(p
{\overline p}\rightarrow {\overline t} t h)$ on
the renormalization/factorization scale $\mu$, at $\sqrt{s}=14~TeV$,
 for $M_h=120~GeV$.
 (b) $\sigma_{NLO}$ and $\sigma_{LO}$ for $p {\overline p}
\rightarrow t {\overline t} h$ as a function of $M_h$, at 
$\sqrt{s}=2~TeV$, for $\mu=M_t$ and $\mu=2 M_t$. From Ref [15].}
\label{lhc_tth}
\end{figure*}

\section{Tevatron}
Higgs boson production rates at the Tevatron are much smaller than at the LHC,
but with the increased luminosity of Run II, it may be possible to
observe a Higgs signal for a Higgs mass below around
 $180~GeV$\cite{tevrun2}.
The dominant production mode is gluon fusion, with a cross section
between $1.0$ and $0.2~pb$ at $\sqrt{s}=2~TeV$ for $M_h$ in
the $120-180~GeV$ region.     Gluon fusion, however,
suffers from large QCD backgrounds to the
dominant $h\rightarrow b {\overline b} $ decay channel.
The gluon fusion production mechanism may, however,  be useful for 
$140 < M_h < 180~GeV$, when combined with the $h\rightarrow W W^*$
decay channel. 

\begin{figure}[t]
\begin{center}
\vspace*{.8cm}
\hspace*{-1.5cm}
\epsfxsize=9cm \epsfbox{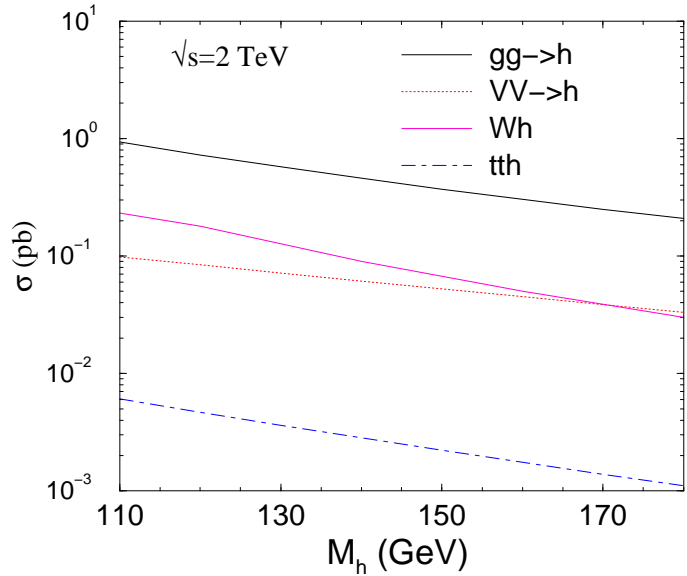}
\caption[ ]{NLO rates for Higgs production in $p {\overline p}$
collisions
at $\sqrt{s}=2~TeV$, evaluated at the renormalization/factorization
scale $\mu=M_h$. (The process $p {\overline p} \rightarrow t {\overline
t} h$ is evaluated at $\mu=2 m_t$).}
\label{tevrates}
\end{center}
\end{figure}

 The most likely discovery
 channel at the Tevatron for $M_h< 140~GeV$ is
the associated production of $Wh$ or $Zh$, where an efficient
trigger is provided by
the leptonic decay modes of the vector bosons.   For Higgs bosons in
the $M_h\sim 120~GeV$ region, the associated production with a top
quark pair may also be observable.\cite{rain}  The status of
the  NLO QCD corrections to these
processes is  briefly discussed below and the rates at the Tevatron,
 including NLO QCD corrections for all channels, are shown in Fig.
\ref{tevrates}. (The rate for $Zh$ production is about a factor of $2$ below
that for $Wh$ production).

\subsection{Gluon Fusion}

At the Tevatron, gluon fusion contributes roughly $65\%$ of the total
Higgs production cross section for $120~GeV< M_h < 180~GeV$. 
The NLO corrections to inclusive Higgs production 
 are large and positive for all values of the Higgs mass.  
The NNLO corrections to $gg\rightarrow h$ at the Tevatron have
been computed by Harlander and Kilgore\cite{hk} and by Catani
${\it et. al.}$\cite{catani} 
in the soft plus collinear approximation
described above 
and are shown in
Fig. \ref{kfac_hk}.  Given the large numerical value of these
partial  corrections, a
complete NNLO calculation is essential before reliable predictions
can be made in this channel.

\subsection{Associated Production, $p {\overline p}\rightarrow
Wh,~Zh$}

The $Wh$ and $Zh$ channels are the most promising discovery channels
  at the Tevatron for $M_h < 140~GeV$.  The
NLO rate for $p {\overline p}\rightarrow Wh$ is shown in Fig. \ref{tevrates}
and is around $.1-.2~pb$. (This figure does not include the $W$ and $h$
decay branching ratios.) The
NLO QCD corrections are the same as those for Drell-Yan  and increase
the rate by about $30\%$ from the lowest order prediction.\cite{wh}
The dependence of the NLO corrected rate on the choice of parton 
distribution functions is quite small, but there remains about
a $12\%$ uncertainty in the prediction due to the residual 
renormalization/factorization scale dependence.

\begin{figure}[t]
\begin{center}
\vspace*{.8cm}
\hspace*{-1.5cm}
\epsfxsize=9cm \epsfbox{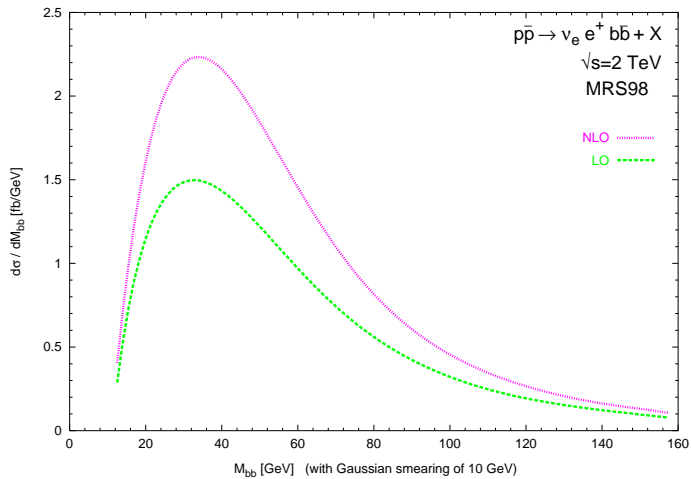}
\caption{The $m_{b{\bar b}}$ distribution of the
$W^+(\to \nu_e e^+) \, b{\bar b}$
background to the Higgs signal, ($p {\overline p}
\rightarrow Wh$), at LO and NLO at $\sqrt{s}=2~TeV$.
 The $b$-quark tagging
efficiency is not included.  From Ref. [19].}
\label{fig:wbbmbb}
\end{center}
\end{figure}

Since the dominant decay of a Higgs boson below $M_h\sim 140~GeV$ is
to $b {\overline b}$ pairs, 
the irreducible background processes to $p {\overline p}\rightarrow Wh$
and $p {\overline p}\rightarrow Zh$ are
$p {\overline p} \rightarrow W b {\overline b}$ and 
$p {\overline p}\rightarrow Z b {\overline b}$.
These background processes 
 have been calculated to NLO in Ref.\cite{ce} and the
results implemented in the Monte Carlo program, MCFM.
The NLO corrections are large and positive and change the
shape of the $b {\overline b}$ distribution near the peak,
as can be seen in Fig. \ref{fig:wbbmbb}.  The $K$
factors for the background $W b {\overline b}$ and $Z b {\overline b}$
processes are larger than those for the $Wh$ and $Zh$ signals
 and have not 
been included in the studies of Ref. \cite{ce}.

\subsection{Associated Production, $p {\overline p}\rightarrow t
{\overline t} h$}
At the Tevatron, the associated production of  $t {\overline t}h $
proceeds primarily through $ q {\overline q}
$ annihilation.  While the rate for $t {\overline t} h$ production is 
small, the signature 
is  distinctive. Unlike at the LHC, at the Tevatron the 
invariant mass distributions
of the final state $b {\overline b}$ pairs from the
$t {\overline t}h$ signal
 have rather a different shapes from the background
and preliminary studies suggest that it may be possible to observe
this channel at the Tevatron.\cite{rain}

   The next to leading order results have been
computed recently by two groups, with excellent agreement.\cite{rd,ben}
The NLO result shows a reduced scale dependence
from the lowest order result
 and a slightly reduced cross section from the lowest order prediction.
For example, for $M_h=120~GeV$ and $\mu=M_t$, the NLO total 
cross section is reduced to $4.86\pm0.03~fb$ from the lowest
order prediction of $6.868\pm.002~fb$.  The reduction
is much less dramatic at $\mu=2 M_t$, as can be seen from Fig.
\ref{tev_lonlo_fig}.
Only for renormalization/factorization scales larger
around than $\mu=2M_t+M_h$ is the NLO cross section larger than the
lowest order rate.
Combining the residual scale dependence with the error from the 
parton distribution functions $(\sim 6\%)$ and
from $m_t$ ($7\%$), we estimate the uncertainty on the
theoretical prediction as about $12\%$.

\newdimen\figwid
\figwid=\columnwidth
\multiply\figwid by 4
\divide\figwid by 9
\begin{figure*}
\includegraphics[width=\figwid]{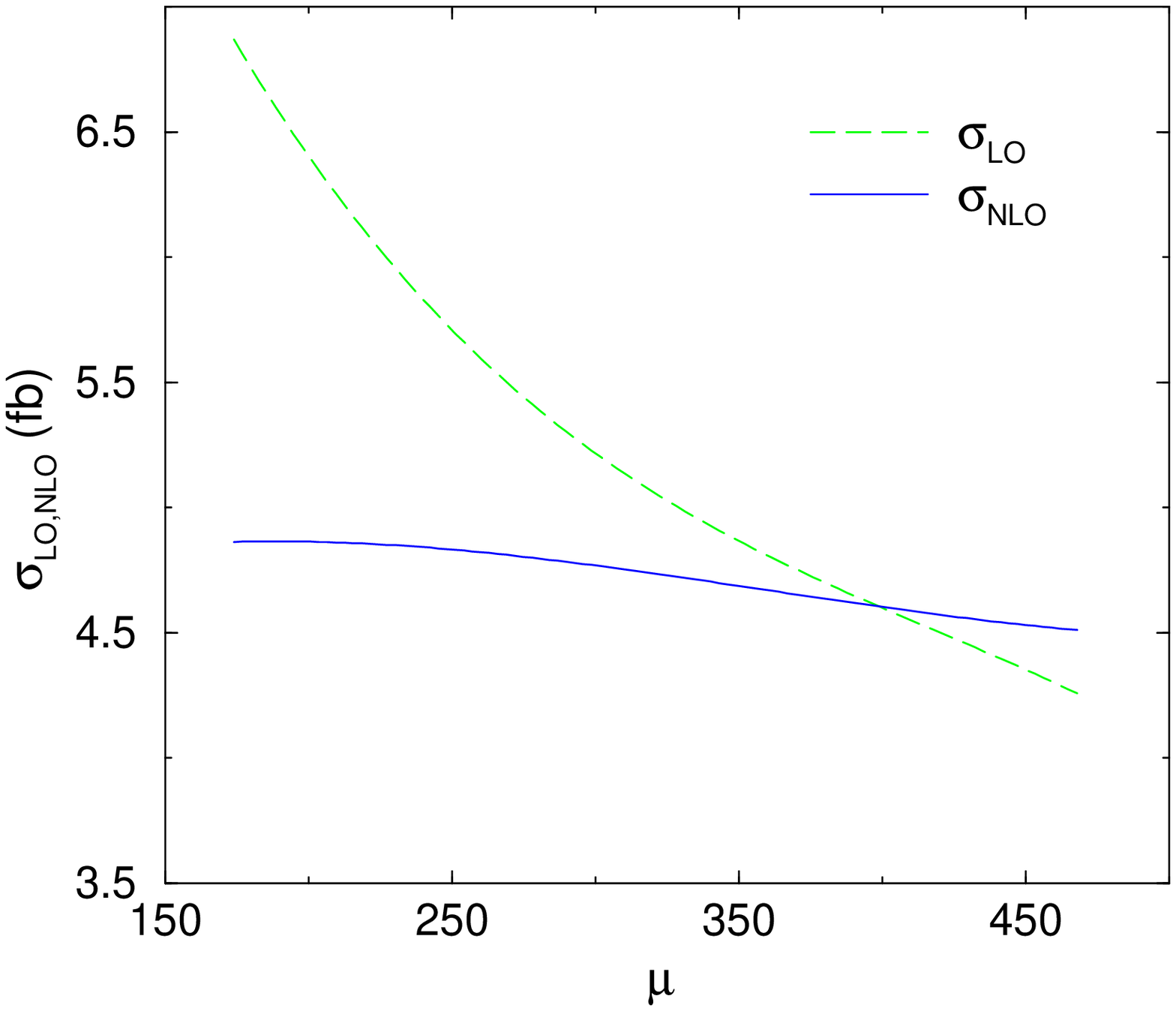}%
\hfil
\includegraphics[width=\figwid]{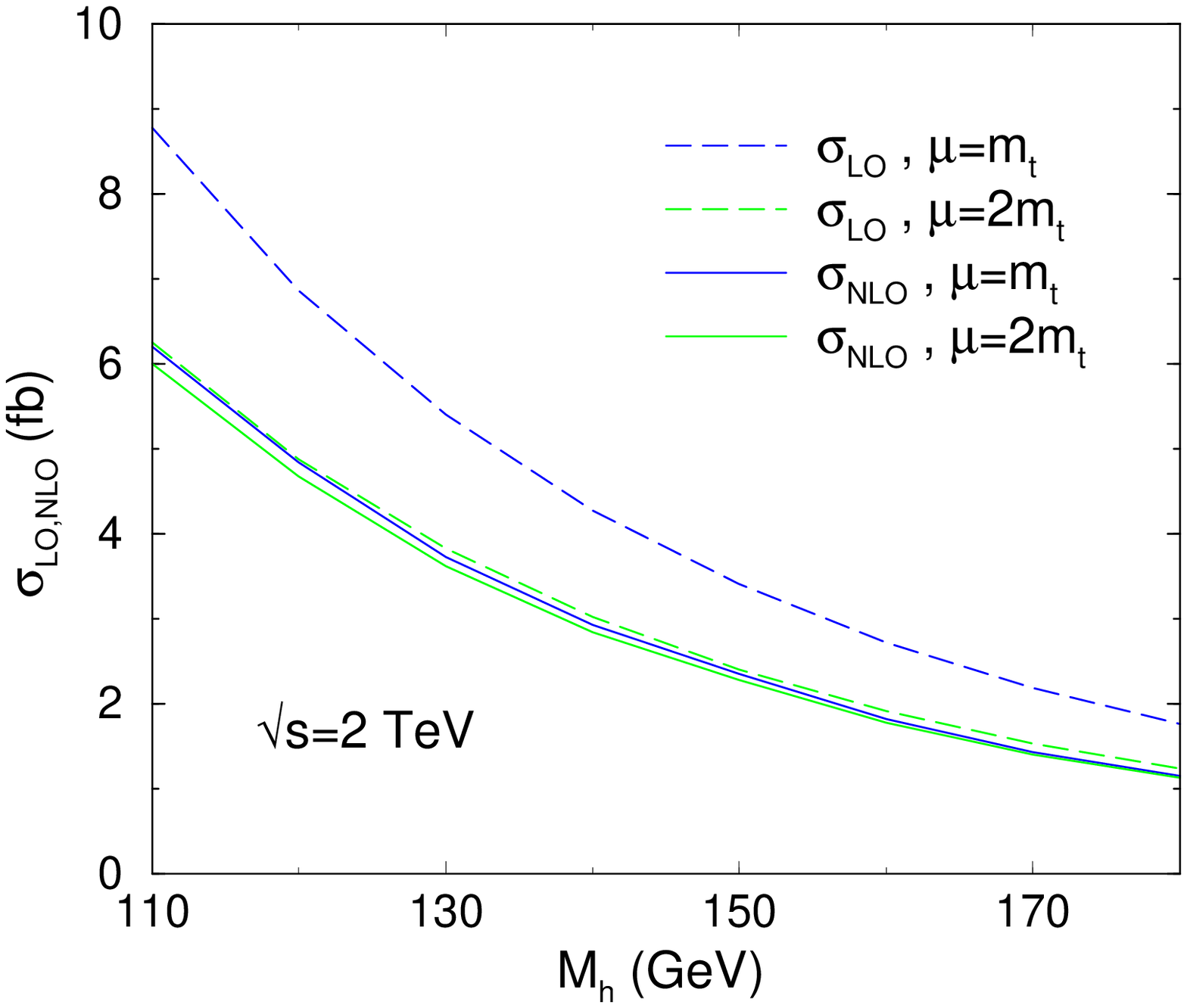}%
\caption{(a) Dependence of $\sigma_{LO, NLO}(p
{\overline p}\rightarrow {\overline t} t h)$ on
the renormalization/factorization scale $\mu$, at $\sqrt{s}=2~TeV$,
 for $M_h=120~GeV$.
 (b) $\sigma_{NLO}$ and $\sigma_{LO}$ for $p {\overline p}
\rightarrow t {\overline t} h$ as a function of $M_h$, at 
$\sqrt{s}=2~TeV$ for $\mu=M_t$ and $\mu=2 M_t$. From Ref [20].}
\label{tev_lonlo_fig}
\end{figure*}

\section{Conclusions}
NLO corrections to all Higgs production channels of interest at hadron 
colliders have now been completed. These NLO predictions show a 
significantly reduced renormalization/factorization scale 
dependence from the LO predictions, leading to increased confidence
in the validity of the predictions.   Complete
 NNLO corrected rates for  inclusive Higgs production
should be available soon.  Consistent NNLO calculations
will, however,
 require structure functions derived to NNLO, which are not
yet available.

\begin{acknowledgments}
I thank my collaborators, Laura Reina and Doreen Wackeroth,
for many valuable discussions.  This work is supported by the U.S. Department
of Energy under grant DE-AC02-76CH00016.
\end{acknowledgments}


\end{document}